\documentclass[equations,11pt]{article}
\usepackage{amssymb,amsmath,amsthm}
\usepackage{eucal}
\usepackage{epsfig,epic,picinpar}

\renewcommand{\theequation}{\arabic{equation}}
\begin{document}
\bibliographystyle{plain}
\def\m@th{\mathsurround=0pt}
\mathchardef\bracell="0365 
\def\upbrall{$\m@th\bracell$}
\def\undertilde#1{\mathop{\vtop{\ialign{##\crcr
    $\hfil\displaystyle{#1}\hfil$\crcr
     \noalign
     {\kern1.5pt\nointerlineskip}
     \upbrall\crcr\noalign{\kern1pt
   }}}}\limits}
\mathchardef\hatcell="0352 
\def\dobrall{$\m@th\hatcell$}
\def\underhat#1{\mathop{\vtop{\ialign{##\crcr
    $\hfil\displaystyle{#1}\hfil$\crcr
     \noalign
     {\kern1.5pt\nointerlineskip}
     \dobrall\crcr\noalign{\kern1pt
   }}}}\limits}
\def\theequation{\arabic{section}.\arabic{equation}}
\newcommand{\ar}{\alpha}
\newcommand{\aar}{\bar{a}}
\newcommand{\bb}{\beta}
\newcommand{\gm}{\gamma}
\newcommand{\Gm}{\Gamma}
\newcommand{\en}{\epsilon}
\newcommand{\ven}{\varepsilon}
\newcommand{\dd}{\delta}
\newcommand{\sg}{\sigma}
\newcommand{\kp}{\kappa}
\newcommand{\ld}{\lambda}
\newcommand{\oa}{\omega}
\newcommand{\be}{\begin{equation}}
\newcommand{\ee}{\end{equation}}
\newcommand{\bea}{\begin{eqnarray}}
\newcommand{\eea}{\end{eqnarray}}
\newcommand{\bse}{\begin{subequations}}
\newcommand{\ese}{\end{subequations}}
\newcommand{\nn}{\nonumber}
\newcommand{\vf}{\varphi}
\newcommand{\sn}{{\rm sn}}
\newcommand{\cn}{{\rm cn}}
\newcommand{\dn}{{\rm dn}}
\newcommand{\wh}{\widehat}
\newcommand{\ol}{\overline}
\newcommand{\wt}{\widetilde}
\newcommand{\ut}{\undertilde}
\newcommand{\uh}{\underhat}
\newcommand{\ip}{{i^\prime}}
\newcommand{\jp}{{j^\prime}}
\newcommand{\btP}{\,^{t\!}{\bf P}}
\newcommand{\bI}{{\bf I}}
\newcommand{\bO}{{\bf O}}
\newcommand{\bA}{{\bf A}}
\newcommand{\bB}{{\bf B}}
\newcommand{\bU}{{\bf U}}
\newcommand{\bC}{{\bf C}}
\newcommand{\bOm}{{\bf \Omega}}
\newcommand{\buk}{{\bf u}_k}
\newcommand{\bul}{{\bf u}_\ell}
\newcommand{\tII}{\,^{t\!}{\bf I}}
\newcommand{\tuk}{\,^{t\!}{\bf u}_{k^\prime}}
\newcommand{\tul}{\,^{t\!}{\bf u}_{\ell^\prime}}
\newcommand{\tull}{\,^{t\!}{\bf u}_{-\ell+\ld}}
\newcommand{\tuq}{\,^{t\!}{\bf u}_{-q_j+\ld}}
\newcommand{\tck}{\,^{t\!}{\bf c}_{k^\prime}}
\newcommand{\ssk}{\sigma_{k^\prime}}
\newcommand{\ssl}{\sigma_{\ell^\prime}}
\newcommand{\pte}{(\partial_t-\partial_\eta)}  
\newcommand{\pxe}{(\partial_x-\partial_\eta)}  
\newcommand{\dint}{\int_\Gamma d\mu(\ell) }
\def\hypotilde#1#2{\vrule depth #1 pt width 0pt{\smash{{\mathop{#2}
\limits_{\displaystyle\widetilde{}}}}}}
\def\hypohat#1#2{\vrule depth #1 pt width 0pt{\smash{{\mathop{#2}
\limits_{\displaystyle\widehat{}}}}}}
\def\hypo#1#2{\vrule depth #1 pt width 0pt{\smash{{\mathop{#2}
\limits_{\displaystyle{}}}}}}

\newcommand{\pii}{P$_{{\rm\small II}}$}  
\newcommand{\pvi}{P$_{{\rm\small VI}}$}   
\newcommand{\ptf}{P$_{{\rm\small XXXIV}}$}   
\newcommand{\diffE}{O$\triangle$E}   

\newcommand{\pp}{\partial}
\newcommand{\hf}{\frac{1}{2}}
\newcommand{\ith}{$i^{\rm th}$\ }
\newcommand{\bu}{{\boldsymbol u}}
\newcommand{\bell}{{\boldsymbol l}}
\newcommand{\bj}{{\boldsymbol j}}
\newcommand{\bt}{{\boldsymbol t}}
\newcommand{\bm}{{\boldsymbol m}}
\newcommand{\boa}{{\boldsymbol \omega}}
\newcommand{\bet}{{\boldsymbol \eta}}
\newcommand{\bW}{\bar{W}}
 \newcommand{\pl}{\partial}
 \newcommand{\ddp}{\frac{\partial}{\partial p}}
 \newcommand{\ddq}{\frac{\partial}{\partial q}}
 \newcommand{\ddr}{\frac{\partial}{\partial r}}
 \newcommand{\Ld}{{\boldsymbol \Lambda}}
 \newcommand{\tLd}{\,^{t\!}{\boldsymbol \Lambda}}
 \newcommand{\I}{{\bf I}}
 \newcommand{\bP}{{\bf P}}
 \newcommand{\tbP}{\,^{t\!}{\bf P}}
 \newcommand{\tbC}{\,^{t\!}{\bf C}}
 \newcommand{\ddint}{\int_\Gamma d\ld(\ell) }
 \newcommand{\vE}{\vec{E} }
 \newcommand{\vL}{\vec{L} }
 \newcommand{\vn}{\vec{n} }
 \newcommand{\vR}{\vec{R} }
 \newcommand{\vP}{\vec{P} }
 \newcommand{\vna}{\vec{\nabla} }
 \newcommand{\vv}{\vec{v} }
 \newcommand{\vF}{\vec{F} }
 \newcommand{\vj}{\vec{j} }
 \newcommand{\vB}{\vec{B} }
 \newcommand{\vr}{\vec{r} }
 \newcommand{\vp}{\vec{p} }
 \newcommand{\vk}{\vec{k} }
\newcommand{\mbe}{{\boldsymbol e}}
\newcommand{\bE}{{\boldsymbol E}}
\newcommand{\bnab}{{\boldsymbol \nabla}}
\newcommand{\buu}{{\boldsymbol u}}
\newcommand{\bv}{{\boldsymbol v}}
\newcommand{\ba}{{\boldsymbol a}}
\newcommand{\bbb}{{\boldsymbol b}}
\newcommand{\bS}{{\boldsymbol S}}
\newcommand{\bT}{{\boldsymbol T}}
\newcommand{\bJ}{{\boldsymbol J}}
\newcommand{\bc}{{\boldsymbol c}}
\newcommand{\bw}{{\boldsymbol w}}
\newcommand{\mbx}{{\boldsymbol x}}
\newcommand{\mby}{{\boldsymbol y}}
\newcommand{\bz}{{\boldsymbol z}}
\newcommand{\brr}{{\boldsymbol r}}
\newcommand{\bp}{{\boldsymbol p}}
\newcommand{\bk}{{\boldsymbol k}}
\newcommand{\btt}{{\boldsymbol t}}
\newcommand{\bmm}{{\boldsymbol m}}
\newcommand{\bdd}{{\boldsymbol \delta}}
\newcommand{\bze}{{\boldsymbol 0}}
\newcommand{\boma}{{\boldsymbol \omega}}
\newcommand{\bxi}{{\boldsymbol \xi}}
 \newcommand{\mbv}{\boldmath{v}}
 \newcommand{\mbxi}{\boldmath{\xi}}
 \newcommand{\mbeta}{\boldmath{\eta}}
 \newcommand{\mbw}{\boldmath{w}}
 \newcommand{\mbu}{\boldmath{u}}

 \def\hypotilde#1#2{\vrule depth #1 pt width 0pt{\smash{{\mathop{#2}
 \limits_{\displaystyle\widetilde{}}}}}}
 \def\hypohat#1#2{\vrule depth #1 pt width 0pt{\smash{{\mathop{#2}
 \limits_{\displaystyle\widehat{}}}}}}
 \def\hypo#1#2{\vrule depth #1 pt width 0pt{\smash{{\mathop{#2}
 \limits_{\displaystyle{}}}}}}

\newtheorem{theorem}{Theorem}[section]
\newtheorem{lemma}{Lemma}[section]
\newtheorem{cor}{Corollary}[section]
\newtheorem{prop}{Proposition}[section]
\newtheorem{definition}{Definition}[section]
\newtheorem{conj}{Conjecture}[section]

\begin{flushright}
\end{flushright}
\begin{center} 
{\large{\bf Lax pair for the Adler (lattice Krichever-Novikov) System}} 
\vspace{.4cm} 

F.W. Nijhoff\\ 
{\it Department of Applied Mathematical Studies \\
University of Leeds, Leeds LS2 9JT, U.K.}
\vspace{.2cm}

\end{center} 
\vspace{.4cm}
\centerline{\bf Abstract}
\vspace{.2cm}

\noindent 
In the paper [V. Adler, IMRN {\bf 1} (1998) 1--4] a lattice version 
of the Krichever-Novikov equation was constructed. We present in this 
note its Lax pair and discuss its elliptic form.

\pagebreak

\section{Introduction} 
\setcounter{equation}{0} 

Integrable nonlinear evolution equations that can be solved in
terms of linear problems with spectral parameter on an elliptic curve 
have been known since a number of years, the most well-known example 
being the Landau-Lifschitz (LL) equation, cf. \cite{Skly}, i.e. 
\begin{equation} \label{eq:LL} 
\bS_t=\bS\times\bS_{xx}+\bS\times\bJ\bS \    . 
\end{equation}  
Here $\bS$ is the normalised spin-vector $\bS\in\mathbb R^3$, and 
$\bJ={\rm diag}(J_1,J_2,J_3)$ is the diagonal matrix of anisotropy 
parameters. 

Another well-known example is the Krichever-Novikov (KN) equation which 
reads 
\begin{equation} \label{eq:KN} 
\xi_t=\frac{1}{4}\xi_{xxx}+\frac{3(1-\xi_{xx}^2)}{ 8\xi_x} -\frac{3}{2}
\wp(2\xi)\xi_x^3\    ,  
\end{equation}  
with $\wp$ being the Weierstrass $\wp$-function. 
Eq. (\ref{eq:KN}) arose in \cite{KN} from the general problem 
of describing finite-gap solutions of the Kadomtsev-Petviashvili 
(KP) equation associated with commuting differential operators 
in the KP hierarchy. Eq. (\ref{eq:KN}) applies to the particular case 
that these operators describe a subalgebra generated by second order 
differential operators, i.e. the common eigenspaces are of dimension 2 
(i.e. the ``rank 2'' situation) and obey an algebraic relation 
which determines an elliptic curve, cf. also \cite{Grun1}. 

In its equivalent rational form the equation reads 
\begin{equation}\label{eq:KNrat} 
u_t=\frac{1}{4}u_{xxx}+\frac{3}{8}\frac{r(u)-u_{xx}^2}{u_x}\   , 
\end{equation}  
for $u\equiv\wp(\xi)$ and where $r(u)=4u^3-g_2u-g_3$ is the polynomial 
of the standard Weierstrass curve 
\begin{equation} \label{eq:Weier}  
\Gamma:~~~U^2=r(u)=4u^3-g_2u-g_3  
\end{equation} 
(in principle by performing homographic transformations of the type 
\[ u~~\rightarrow~~ \frac{\ar u+\bb}{\gm u +\dd}  \] 
$r(u)$ can be replaced by a general quartic 
polynomial ). 
Degenerate cases (when the curve (\ref{eq:Weier}) 
reduces to a rational curve) can be mapped to the Schwarzian KdV (SKdV) equation 
\begin{equation}\label{eq:SKdV} 
\frac{z_t}{z_x}=\frac{1}{4}\{\,z\,,\,x\,\}~~~,~~~ 
\{\,z\,,\,x\,\}\equiv \frac{z_{xxx}}{z_x}-\frac{3}{2}\frac{z_{xx}^2}{z_x^2} ~~, 
\end{equation}  
the brackets denoting the Schwarzian derivative. The latter equation has a 
beautiful discrete counterpart on the 2-dimensional lattice, namely 
\begin{equation} \label{eq:dSKdV} 
\frac{(z_{n,m}-z_{n+1,m})(z_{n,m+1}-z_{n+1,m+1})}
{(z_{n,m}-z_{n,m+1})(z_{n+1,m}-z_{n+1,m+1})}=\frac{q^2}{p^2}
\end{equation} 
which (together with its Lax pair) was first given in \cite{NC}, cf. also \cite{Dorf}. 
The r.h.s. of eq. (\ref{eq:dSKdV}) taking the form of the canonical cross-ratio 
of four points in the complex plane, this discrete equation turned out 
to play the defining role in a theory of discrete conformal maps, cf. 
\cite{Bobenko}. It also played an earlier role in the theory of Pad\'e 
Approximants, cf. e.g. \cite{Cordellier}. 

As far as integrable discrete systems associated with elliptic curves 
is concerned, a number of results exists on nonlinear chains, i.e. 
differential-difference equations (D$\Delta$E's), 
mostly due to the group from Ufa, cf. e.g. \cite{Yamil,Shabat}. 
One particular case of such a D$\Delta$E is the following example due 
to Yamilov, cf. \cite{Yamil}, 
\begin{equation}\label{eq:Yamil} 
\frac{du_n}{dt}=(4u_n^3-g_2u_n-g_3)\left( \frac{1}{u_{n+1}-u_n}
+\frac{1}{u_n-u_{n-1}} \right)\   , 
\end{equation}  
which constitutes a differential-difference analogue of the KN 
equation. The elliptic form of this equation is given by:
\begin{equation}
\pl_t\xi_n=\zeta(\xi_n+\xi_{n+1})+\zeta(\xi_n-\xi_{n+1})-
\zeta(\xi_n+\xi_{n-1}) -\zeta(\xi_n-\xi_{n-1})  
\end{equation}  
with $u_n=\wp(\xi_n)$ and where $\zeta$ denotes the Weierstrass 
$\zeta$-function. 
In this context an extension of the theory developed in \cite{KN} 
to commuting difference equations that was recently proposed in \cite{KN2} should be 
mentioned as well.  

On the level of integrable {\it partial difference} equations (P$\Delta$E's), i.e. equations with 
both spatial as well as temporal independent variable discrete, the situation is much less 
developed.  To my knowledge the first 
example of a fully discrete system (discrete in space as well as time) 
was given in \cite{LL}, where a lattice version of the LL equations 
was constructed from an {\it Ansatz} for a Lax pair. 
This resulting lattice system can be retrieved from the basic equations: 
\begin{subequations}\begin{eqnarray}\label{eq:dLL} 
T_0\wh{\bS}-\wh{\bS}\times\bJ\bT&=&\wt{T}_0\bS-\bJ\wt{\bT}\times\bS \\      
S_0\wt{\bT}-\wt{\bT}\times\bJ\bS&=&\wh{S}_0\bS-\bJ\wh{\bS}\times\bT \\ 
\wh{S}_0T_0-\wt{T}_0S_0&=& \ar\left(\wh{\bS}\cdot\bJ^{-1}\bT-
\wt{\bT}\cdot\bJ^{-1}\bS\right)     
\end{eqnarray}  \end{subequations} 
with $\ar=J_1J_2J_3$ and where $S_0$, $T_0$ are scalar fields and the three-vectors $\bS$, $\bT$ 
in $\mathbb R^3$ are normalised to have unit length. How this system 
is related to other discretisations of the LL equations, cf. e.g. 
\cite{Adler2}, is yet unknown. 

In the present letter we investigate a more recent and simpler example of a P$\Delta$E 
associated with elliptic curves that was found by V. Adler in \cite{Adler} on the 
basis of B\"acklund transformations (BT's) for the KN equation. The main point of this 
letter is to present the Lax pair for this lattice system and to exhibit its connection to other 
lattice systems, notably its elliptic form.

\section{The lattice KN System} 
\setcounter{equation}{0} 

A lattice version of the KN equation (\ref{eq:KNrat}) was found 
recently by V. Adler in \cite{Adler} as the permutability condition 
for the BT's of the KN equation. For the purpose of fixing the notations, 
we we briefly recall here its construction. 

The form of the auto-BT mapping solutions of eq. (\ref{eq:KN}) to 
solutions of itself was proposed to be of the following form: 
\begin{equation}\label{eq:KNBT} 
u_x\wt{u}_x=\frac{1}{A} H(u,\wt{u},a)  
\end{equation}
with 
\begin{equation}\label{eq:H} 
H(x,y,z)\equiv \left( xy+xz+yz+\frac{g_2}{4}\right)^2 - 
(x+y+z)(4xyz-g_3)   
\end{equation}
and where $(a,A)=(\wp(\ar),\wp'(\ar))$ is a point on the elliptic 
curve $\Gamma$. 

{\bf Remark:} As indicated in \cite{Adler} this result can be generalised 
to include non-auto-B\"acklund transformations between solutions of 
different versions of the KN equations associated with different 
elliptic curves. In fact, if $h(u,v)$ is any biquadratic polynomial, then 
the equation $h=0$ defines a birational correspondence between 
the curves $w^2=r(u)$ and $z^2=R(v)$ obtained from: 
\[ r(u)=h_v^2-2h h_{vv}~~~~~ {\rm and}~~~~~ R(v)=h_u^2-2hh_{uu}\   .   \] 

The permutability condition for the BT, i.e. the condition that 
the BT given by  
\begin{equation}\label{eq:KNBT2} 
u_x\wh{u}_x=\frac{1}{B} H(u,\wh{u},b) 
\end{equation}
(where again $(b,B)=(\wp(\bb),\wp'(\bb))$ is a point on the curve 
$\Gamma$) \textit{commutes} with the original BT leads to 
the algebraic equation of the form: 
\begin{eqnarray}\label{eq:dKN} 
&& k_0 u\wh{u}\wt{u}\wh{\wt{u}}
-k_1\left( u\wh{u}\wt{u}+ u\wh{u}\wh{\wt{u}}+ 
u\wt{u}\wh{\wt{u}}+\wh{u}\wt{u}\wh{\wt{u}}\right) 
+k_2\left( \wh{u}\wt{u}+ u\wh{\wt{u}}\right) \nn \\ 
&&-k_3\left( u\wt{u} +\wh{u}\wh{\wt{u}}\right) 
-k_4\left( u\wh{u} +\wt{u}\wh{\wt{u}}\right) 
+k_5\left( u+\wt{u}+\wh{u}+\wh{\wt{u}}\right)+k_6=0\   .  
\end{eqnarray}
In fact, eliminating the derivatives from (\ref{eq:KNBT}) and 
(\ref{eq:KNBT2}) one obtains the expression:
\[ B^2H(u,\wt{u},a)H(\wh{u},\wh{\wt{u}},a)-
A^2H(u,\wh{u},b)H(\wt{u},\wh{\wt{u}},b)=0  \] 
the l.h.s. of which factorises into two branches, one being the 
l.h.s. of (\ref{eq:dKN}) and the other being a similar expression 
with $B\rightarrow -B$. Thus one obtains the parametrisation of the 
coefficients: 
$$ k_0=A+B~~~,~~~ k_1=aB+bA~~~,~~~ k_2=a^2B+b^2A $$
$$ k_3=\frac{AB(A+B)}{2(b-a)}-b^2A+B\left( 2a^2-\frac{g_2}{4}\right)$$ 
\begin{equation}\label{eq:coeffs} 
k_4=\frac{AB(A+B)}{2(a-b)}-a^2B+A\left( 2b^2-\frac{g_2}{4}\right) 
\end{equation} 
$$ k_5=\frac{g_3}{2}k_0+\frac{g_2}{4}k_1~~~~,~~~~ 
k_6=\frac{g_2^2}{16}k_0+g_3 k_1$$ 
where both $(a,A)$ and $(b,B)$ are points of $\Gamma$, i.e. we have 
the relations ~$A^2=r(a)$, ~$B^2=r(b)$~. 

In eq. (\ref{eq:dKN}) we have adopted our preferred short-hand notation 
for lattice systems with $u=u_{n,m}$, $\wt{u}=u_{n+1,m}$, 
$\wh{u}=u_{n,m+1}$ and $\wh{\wt{u}}=u_{n+1,m+1}$ denoting the values of 
the dependent variable $u$ around an elementary plaquette, cf. 
Figure 1. The $\alpha$, $\beta$ in Figure 1 are lattice parameters 
representing the grid size. Alternatively, we can think of the lattice 
parameters in the case of (\ref{eq:dKN}) to take values 
as points on the elliptic curve (\ref{eq:Weier}), i.e. they are 
parametrised by $\ar$ and $\bb$ as ~
$$(a,A)=\left(\wp(\ar),\wp'(\ar)\right)~~~~,~~~~ 
(b,B)=\left(\wp(\bb),\wp'(\bb)\right)\  . $$ 
\vspace{.3cm} 

\setlength{\unitlength}{1mm}
\begin{picture}(40,40)(-60,0) 

\put(0,0){\circle*{3}}
\put(0,30){\circle*{3}}
\put(30,30){\circle*{3}}
\put(30,0){\circle*{3}}

\put(0,0){\vector(1,0){15}}
\put(15,0){\vector(1,0){15}}
\put(0,30){\vector(0,-1){15}}
\put(0,15){\vector(0,-1){15}}
\put(0,30){\vector(1,0){15}}
\put(15,30){\vector(1,0){15}}
\put(30,30){\vector(0,-1){15}}
\put(30,15){\vector(0,-1){15}}

\put(-5,-5){\large$\wh{u}$}
\put(34,-5){$\wh{\wt{u}}$}
\put(34,34){$\wt{u}$}
\put(-5,34){$ u$}

\put(15,-5){$\alpha$}
\put(15,33){$\alpha$}
\put(-4,15){$\beta$}
\put(34,15){$\beta$}

\end{picture} 
\vspace{.3cm} 

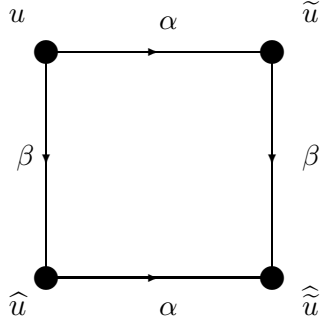
{\bf Figure 1:} {\it Configuration of lattice points in the lattice equation of 
KdV type.} 
\vspace{.2cm} 

We mention that Adler's lattice system (\ref{eq:dKN}) 
can be written in the following remarkable form:  
\begin{eqnarray}
&&A\left[ (u-b)(\wh{u}-b)-(a-b)(c-b)\right] 
\left[(\wt{u}-b)(\wh{\wt{u}}-b) -(a-b)(c-b)\right] \nn \\ 
&&+B\left[ (u-a)(\wt{u}-a)-(b-a)(c-a)\right] 
\left[ (\wh{u}-a)(\wh{\wt{u}}-a) -(b-a)(c-a)\right]= \nn \\ 
&& = ABC(a-b)   \label{eq:ellform} 
\end{eqnarray}
in which $c=\wp(\bb-\ar)$, $C=\wp'(\bb-\ar)$. To obtain (\ref{eq:ellform}) 
we note the following relations among the coefficients $k_i=k_i(\ar,\bb)$ in 
(\ref{eq:coeffs}) 
\[ k_3+k_2=B(a-b)(a-c) ~~~~,~~~~ 
k_4+k_2=A(b-a)(b-c) \  , \] 
as well as 
\begin{eqnarray*}
k_5&=&a(k_2+k_3)+b(k_2+k_4)-Ab^3-Ba^3\  , \\ 
k_6&=&A\left(b^2-(a-b)(c-b)\right)^2+B\left(a^2-(b-a)(c-a)\right)^2 \\  
&& +AB\left[ A(b-c)+B(a-c) \right] \  , 
\end{eqnarray*}
eliminating the moduli $g_2$, $g_3$. Use is made of the addition formulae 
for the Weierstrass $\wp$-function in the form: 
\[ A(c-b)+B(c-a)=C(a-b)\  . \] 

We explain in the next section in what sense we consider an equation 
of the type (\ref{eq:dKN}) to be integrable, and present a derivation of its 
Lax pair.

\section{Integrability of the Adler system} 
\setcounter{equation}{0}

We will here recollect a few general points regarding lattice systems 
of the type presented in the previous section. 
The points made here are not new and were presented and illustrated 
for different lattice systems at numerous occasions, cf. e.g. 
\cite{NW,Finland,Newton}. 

Eq. (\ref{eq:dKN}) is an example in ``general position'' of a large class of 
partial difference equations (P$\Delta$E's) of the following canonical form: 
\begin{equation}\label{eq:geneq} 
f(u,\wt{u},\wh{u},\wh{\wt{u}};\ar,\bb)=0\  ,  
\end{equation}
where we adopt the notation of the previous section to indicate the vertices around 
an elementary plaquette on a rectangular lattice: 
$$ u:=u_{n,m}~~~~~,~~~~~ \wt{u}=u_{n+1,m} $$
$$ \wh{u}:=u_{n,m+1}~~~~~,~~~~ \wt{u}=u_{n+1,m+1} $$
cf. Figure 1. In eq. (\ref{eq:geneq}) $\ar$, $\bb$ denote (as in the previous section) 
lattice parameters, i.e. parameters that are singled out from other (fixed)  
parameters by being associated with the lattice shifts 
\begin{center}
$u~~\stackrel{\ar}{\rightarrow}~~ \wt{u}$ \\ 
$u~~\stackrel{\bb}{\rightarrow}~~ \wh{u}$ 
\end{center}
In fact, we will assume in what follows that the lattice equations (\ref{eq:geneq}) are 
{\it covariant} with respect to the simultaneous interchangement of both 
lattice variables as well as lattice parameters.  As is well-known 
{\it integrable} lattice equations of the type above can be reinterpreted as 
the permutability condition of B\"acklund transformations, 
identifying the lattice parameters as the coresponding 
B\"acklund parameters. Furthermore, the dependence on these parameters 
can be exploited to obtain continuum limits of the lattice equations to 
semi-continuous and fully continuous equations that are compatible 
with the original fully discrete equation, cf. \cite{NQC,QNCL}.  

Examples of (\ref{eq:geneq}) have been studied since many years, cf. 
\cite{NC} for a review. Two important questions that arise in considering 
integrable cases of such equations are the following: 
\begin{enumerate}
\item When does a P$\Delta$E of the form (\ref{eq:geneq}) make sense from 
the point of view of initial value problems (IVP's)? This is the question on 
the well-posedness of the IVP on the lattice.  
\item What is the characteristic property of such equations associated 
with its integrability? We have in mind here the analogous property to the 
existence of conservation laws and higher symmetries. 
\end{enumerate}
Both questions have been answered (for canonical examples) in recent years. The first 
question was answered in a series of papers a decade ago, cf. e.g. \cite{PNC,CNP}. 
Clearly, if the equation $f=0$ can be solved uniquely at each 
vertex of the plaquette (i.e. if it is linear in each of the dependent variables 
associated with its vertices), we can define initial value problems 
on configurations like: 

\setlength{\unitlength}{1mm}
\begin{picture}(40,40)(-30,0) 

\put(-15,15){\circle*{3}}
\put(0,0){\circle*{3}}
\put(15,15){\circle*{3}}
\put(30,0){\circle*{3}}
\put(45,15){\circle*{3}}
\put(60,0){\circle*{3}}
\put(75,15){\circle*{3}}
\put(90,0){\circle*{3}}
\put(105,15){\circle*{3}}

\put(-15,15){\line(1,-1){15}}
\put(0,0){\line(1,1){15}}
\put(15,15){\vector(1,-1){15}}
\put(30,0){\vector(1,1){15}}
\put(45,15){\vector(1,-1){15}}
\put(60,0){\vector(1,1){15}}
\put(75,15){\vector(1,-1){15}}
\put(90,0){\vector(1,1){15}}
\end{picture} 

\vspace{.3cm}

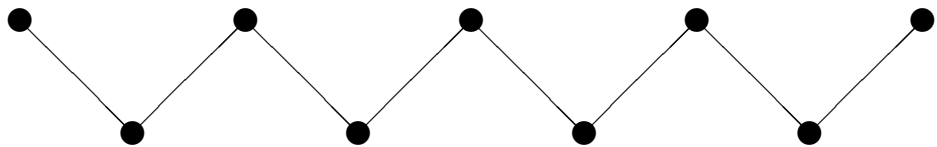
{\bf Figure 2:} {\it Configuration of initial data on the lattice 
for lattice equations of KdV type.} 
\vspace{.2cm} 

The second question, regarding the characteristic of integrability 
of the lattice equations, was discussed in \cite{NW} (as well as in 
various public lectures presented by the author over the past few years). 
There, the main characteristic of the integrability is expressed 
by the fact that in the integrable cases the equation (\ref{eq:geneq}) 
represents actually a {\it parameter-family of consistent equations 
on a multi-dimensional lattice}. It is at this point that the lattice 
parameters $\ar,\bb$ will play a crucial role: instead of fixing them 
once an forever and solving the equation for specific values of the 
parameters, we leave them free and associate new lattice directions 
with each new value of the corresponding lattice parameter. The fact that 
this can actually be done in a consistent way, imposing a multitude 
of equations of the same form (but with different parameters and in different 
latice directions) on one and the same dependent variable $u$ is the 
true characteristic of the integrability of the lattice equation: it is 
a purely combinatorial property of the lattice system that forms the precise 
analogue of the existence of higher symmetries of continuous evolution 
equations of KdV type. 

\paragraph{Proposition:} The lattice equation (\ref{eq:dKN}),  
with the parametrisation of coefficients given by (\ref{eq:coeffs}), 
represents a \textit{compatible parameter-family of partial 
difference equations}, i.e. the equation can be embedded in a consistent 
way into a multidimensional lattice, on each twodimensional sublattice 
of shifts of which an equation of the form (\ref{eq:dKN}) holds for one and 
the same independent variable $u$. 

Let us explain how the consistency of this embedding manifests itself, cf. \cite{NW}. 
As stated above, with each value of the lattice parameters $\ar,\bb$ we can 
associate a discrete variable corresponding to a direction in 
a multidimensional lattice, s.t. the solution $u$ can be considered 
as a \textit{function} on this multidimensional lattice, i.e. a single-valued  
object 
$$u=u_{n,m,h,\dots}=u(n,m,h,\dots;\ar,\bb,\gm,\dots)$$ 
with the notation for the lattice shifts 
$$\wt{u}:=u_{n+1,m,h,\dots}~~~,~~~ \wh{u}:=u_{n,m+1,h,\dots}~~~,~~~ 
\ol{u}:=u_{n,m,h+1,\dots}$$ 

It suffices to investigate how the embedding the lattice equation works in a three-dimensional 
lattice. Thus we impose the system of three equations: 
\bse\label{eq:lsys}\begin{eqnarray}
&& f(u,\wt{u},\wh{u},\wh{\wt{u}};\ar,\bb)=0 \label{eq:lsysa}\\ 
&& f(u,\wt{u},\ol{u},\wt{\ol{u}};\ar,\gm)=0 \label{eq:lsysb}\\ 
&& f(u,\wh{u},\ol{u},\wh{\ol{u}};\bb,\gm)=0 \label{eq:lsysc}  \   ,  
\end{eqnarray}\ese 
and investigate the iteration scheme around an elementary cube for given 
initial values at $u$, $\wt{u}$, $\wh{u}$ and $\ol{u}$ (see Figure 3). 
Iterating along the faces of the cube by using the three equations 
(\ref{eq:lsysa}), (\ref{eq:lsysb}) and (\ref{eq:lsysc}) we arrive at 
a possible point of conflict for the value of $\wh{\wt{\ol{u}}}$ which, in 
principle, can be calculated in three different ways. If, generically, 
these three different ways give the same answer, i.e. if the evaluation 
of the iteration scheme at that corner of the cube is single-valued, the 
system is consistent. It is in this case that we consider the original 
lattice equation to be integrable. 
\vspace{.3cm}

\begin{center}

\setlength{\unitlength}{0.001in}
\begingroup\makeatletter\ifx\SetFigFont\undefined%
\gdef\SetFigFont#1#2#3#4#5{%
  \reset@font\fontsize{#1}{#2pt}%
  \fontfamily{#3}\fontseries{#4}\fontshape{#5}%
  \selectfont}%
\fi\endgroup%
{\renewcommand{\dashlinestretch}{30}
\begin{picture}(3482,2813)(0,-10)
\put(450,1883){\circle*{150}}
\put(75,1883){\makebox(0,0)[lb]{$\ol{u}$}} 

\put(1275,2708){\circle*{150}}
\put(825,2708){\makebox(0,0)[lb]{$u$}} 

\put(3075,2708){\circle*{150}}
\put(3375,2633){\makebox(0,0)[lb]{$\wt{u}$}} 

\put(2250,83){\circle{150}}
\put(2550,8){\makebox(0,0)[lb]{$\wh{\wt{\ol{u}}}$}} 

\put(1275,908){\circle*{150}}
\put(825,908){\makebox(0,0)[lb]{$\wh{u}$}} 

\put(2250,1883){\circle{150}}
\put(1950,2108){\makebox(0,0)[lb]{$\wt{\ol{u}}$}} 

\put(450,83){\circle{150}}
\put(0,8){\makebox(0,0)[lb]{$\wh{\ol{u}}$}} 

\put(3075,908){\circle{150}}
\put(3300,833){\makebox(0,0)[lb]{$\wh{\wt{u}}$}} 

\drawline(1275,2708)(3075,2708)
\drawline(1275,2708)(450,1883)
\drawline(450,1883)(450,83)
\drawline(3075,2708)(2250,1883)
\drawline(450,1883)(2250,1883)
\drawline(3075,2633)(3075,908)
\drawline(1275,908)(450,83)
\drawline(1275,908)(3075,908)
\drawline(2250,1883)(2250,83)
\drawline(450,83)(2250,83)
\drawline(3075,908)(2250,83)
\dashline{60.000}(1275,2633)(1275,908)
\end{picture}
}
\vspace{.3cm}

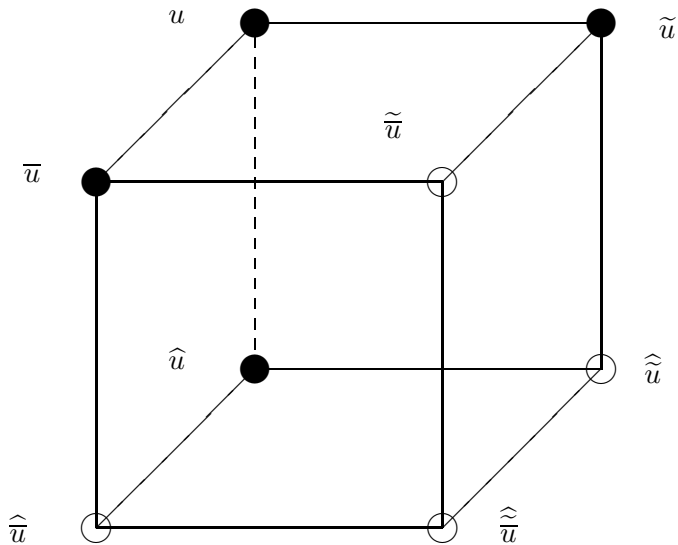
{\bf Figure 3:} {\it Consistency of the lattice system (\ref{eq:dKN}) on the three-dimensional lattice.} 
\end{center}
\vspace{.2cm} 
For all previously considered lattice equations, e.g. the case 
(\ref{eq:dSKdV}), the consistency in 
the sense explained above can be easily verified by hand, and 
calculating the variable at the outer point of the elementary cube in 
terms of the initial values leads to an expression that is invariant 
under the interchange of lattice directions, cf. \cite{W} for details, 
(in the paper \cite{NW} we gave the lattice modified KdV equation as an 
example, where the scheme works precisely in the same way). 
We mention that in a recent submission, \cite{BS}, the consistency 
around the cube was reiterated in a slightly different context. 
The relevant formulae pertaining to the case of the lattice Schwarzian 
KdV of \cite{NC}, eq. (\ref{eq:dSKdV}), were given in the thesis 
\cite{W}, and have also been recovered in that paper. 

Coming back to the main issue of the present paper, the integrability 
of the Adler system (\ref{eq:dKN}), in the sense of the consistency on 
the multidimensional lattice, is straightforward but technically much 
more involved than the previously considered cases of KdV type. The 
difficulty stems from the algebraic relations between the various 
lattice parameters on the elliptic curve, and consequently the use of 
algebraic manipulation tools becomes inevitable in this case. The 
verification that the Adler system (\ref{eq:dKN}) obeys the 
consistency around the cube has been verified explicitely with the 
help of J. Hietarinta using REDUCE. Thus, in the sense of this notion 
of the existence of commuting discrete flows in the multi-dimensional 
lattice, the Adler system is integrable. 

\paragraph{}
We will now {\it derive} the Lax pair for the Adler system from the 
above considerations. This is achieved by noting that, as an immediate 
consequence of the statement that the integrable lattice equations are 
consistently embedded in a multidimensional lattice, one can state that 
in a sense (viewed as a parameter-family of equations rather than a 
single equation on a 2D lattice) {\it the lattice equation forms its 
own Lax pair}. This point of view was presented in a number of recent 
seminars on the subject, cf. \cite{Finland, Newton}, and was 
implemented explicitly for various cases (including e.g. the case of 
the lattice Boussinesq equation), cf. also the thesis \cite{W} where 
the derivation for the case of the lattice SKdV equation 
(\ref{eq:dSKdV}) was presented.  
Let us outline the steps needed in order to implement this idea; they 
are as follows: 
\begin{enumerate}
\item Fix a ``virtual'' direction on the 3D lattice, e.g. the direction 
associated with a shift denoted by ~$u\rightarrow\ol{u}$~ and 
a lattice parameter $\kp$, with corresponding equations given by 
(\ref{eq:lsysb}) and (\ref{eq:lsysc}).  
\item Regard now the shifted object in the virtual direction $\ol{u}$ 
to be a new quantity, while $\wt{u}$ and $\wh{u}$ represent the 
``physical'' shifts on the original variable, and solve for 
$\wt{\ol{u}}$ and $\wh{\ol{u}}$ 
(which can be done due to the multilinearity of the lattice equation). 
This yields expressions which are fractional linear in $\ol{u}$. 
\item Resolve the resulting ``discrete Riccati'' equations by 
linearisation, i.e. set $\ol{u}=f/g$ and separate into linear eqs. 
for $f$ and $g$, leading to a 2$\times$2 matrix system; this will 
produce the Lax matrices up to a common factor $D$ which needs to be 
specified by a determinantal condition. 
\end{enumerate} 
In this way we obtain in a straightforward way the Lax pair for the 
given lattice equation in the form 
\begin{equation}\label{eq:SKdVLax} 
\left\{ \begin{array}{l} \wt{\vf}=L\vf \\ 
 \wh{\vf}=M\vf \end{array} \right. 
\end{equation} 
with ~$\vf\equiv (f,g)^T$~, and where the auxiliary lattice parameter 
$\kp$ plays now the role of the spectral parameter. Of course, the 
above scheme only applies to the case where the lattice equation can 
be consistently embedded into a multidimensional lattice.  

Thus, having assessed the integrability in the above sense for the Adler system (\ref{eq:dKN}), 
it is an easy exercise to implement the steps 1-3 for this lattice system. This leads to a 
Lax representation of the form (\ref{eq:SKdVLax}) where one part of the Lax pair is given by 
\bse \label{eq:dKNLax} 
\begin{equation}
\wt{\vf}=L_\kp(\wt{u},u;\ar) \vf 
\end{equation} 
with Lax matrix:
\begin{equation}\label{eq:Lax}
 L_\kp=\frac{1}{D}\left(\begin{array}{ccc}
k_1u\wt{u}-k_2\wt{u}+k_4 u-k_5 &,& k_3u\wt{u}-k_5(u+\wt{u})-k_6\\
k_0u\wt{u}-k_1(u+\wt{u})-k_3 &,& -k_1u\wt{u}+k_2 u-k_4\wt{u}+k_5
\end{array}\right)\ .  
\end{equation} \ese 
The coefficients in (\ref{eq:Lax}) can be inferred immediately from the 
form of the lattice equations itself, setting ~$k_i=k_i(\ar,\kp)$~, 
i.e. replacing $\bb$ by the spectral parameter $\kp$ in the formulae 
(\ref{eq:coeffs}) for the $k_i$. 
The only subtlety is the choice of the prefactor 
$D=D(u,\wt{u},\ar,\kp)$ which follows from the determinant of the 
matrix in (\ref{eq:Lax}), which 
is calculated to be: 
\[ (a-k)^2KK'H(u,\wt{u},a) ~~~ {\rm with}~~~ k=\wp(\kp)~~,~~
K=\wp'(\kp)~~,~~ K'=\wp'(\kp-\ar)\  . \] 
Choosing $D$ to be equal to the square root of this expression to 
ensure that $\det(L_\kp)=1$. Taking the other part of the Lax pair in 
the same form apart from the obvious replacements: 
$\wt{\phantom{a}}\rightarrow\wh{\phantom{a}}$, $\ar\rightarrow\bb$, 
we have the compatibility  relations 
\[ L_\kp(\wh{\wt{u}},\wh{u};\ar)L_\kp(\wh{u},u;\bb)=
L_\kp(\wh{\wt{u}},\wt{u};\bb)L_\kp(\wt{u},u;\ar) \] 
which are the Lax equations for Adler system (\ref{eq:dKN}).  

One further exercise is to obtain the Lax pair for the BT 
(\ref{eq:KNBT}) itself. Obviously, we can think of this equation as a 
differential-difference equation representing a continuous commuting 
flow to the fully discrete lattice system (\ref{eq:dKN}). The 
corresponding Lax matrix is obtained in precisely the same way as the 
discrete Lax matrices: solve the Riccati equation associated with the 
virtual shift $\ol{\phantom{a}}$, i.e. eq. (\ref{eq:KNBT}) with $a$ 
replaced by $k$, $A$ by $K$, and $\wt{u}$ by $\ol{u}$, and then making 
the same identifications as before to obtain a 2$\times$2 differential 
matrix system. There is now an additive freedom built into 
the system, which is fixed by imposing that the Lax matrix be 
traceless. Thus, we obtain an additional linear equation given by 
\bse\label{eq:KNLax}
\begin{equation}
\vf_x=U_\kp\vf 
\end{equation}  
with 
\begin{equation}
U_\kp=\frac{1}{Ku_x}\left(\begin{array}{cc}
\frac{1}{2}g_3-(u+k)(uk-\frac{1}{4}g_2) & g_3(u+k)+(uk+\frac{1}{4}g_2)^2\\
-(u-k)^2 & -\frac{1}{2}g_3+(u+k)(uk-\frac{1}{4}g_2)
\end{array}\right) 
\end{equation}  \ese 
which supplements the discrete Lax pair (\ref{eq:dKNLax}). The linear 
equation (\ref{eq:KNLax}) is the spatial part of the Lax pair for the 
continuous KN equation (\ref{eq:KNrat}), which can be recovered from 
the original Lax pair given in the paper \cite{KN}. 

Thus, we have obtained the Lax pair the lattice KN equation 
(\ref{eq:dKN}) of Adler directly from the equation itself. After 
having obtained these results, which were presented during the workshop 
on {\it Discrete Systems and Integrability} at the Isaac Newton 
Institute (september 2001), cf. \cite{Newton}, V. Adler 
communicated to the author that similar (yet unpublished) 
results had been obtained by Yu. Suris and himself, \cite{private}.

\section{Discussion} 
\setcounter{equation}{0} 

In this paper we presented a novel Lax pair for the lattice 
Krichener-Novikov equation of Adler. The derivation is based on the 
simple observation that in a precise sense the lattice equation 
{\it is its own Lax pair}. This is based on the nontrivial fact that 
the lattice equation under consideration represent a consistent 
{\it parameter-family} of equations which are embedded 
in a lattice of arbitrary dimensionality. The ``consistency around a 
cube'', cf. Figure 3,  which demonstrates this property was first 
presented in a paper \cite{NW} and the derivations of Lax pair 
in cases such as the lattice BSQ and SKdV equations was implemented 
in the thesis \cite{W}. In the recent submission \cite{BS} some of 
these results were given a geometrical context. 
equation that were already presented in \cite{W}. 

Although the above observation is very simple, it is also quite deep 
in that it constitutes really the precise anlogue of the existence of 
hierarchies and of infinite sequences of conservation laws 
that are the hallmark of soliton systems. 

Having obtained the Lax pair for the Adler system one can now embark 
on a more systematic study of this new elliptic system, including its 
finite-dimensional and similarity reductions  leading to finite-dimensional 
mappings associated with spectral problems on the torus and (possibly) 
to Painlev\'e-type of equations possessing isomonodromic deformation problems 
on the torus. The first part of such a study is already underway, cf. 
\cite{NP}. 

Let us finish by making some remarks on the elliptic version of some 
of the equations mentioned. It is easily noted that the BT 
(\ref{eq:KNBT}) from which the Adler system originated 
can be cast into elliptic form by going over to dependent variables 
$u=\wp(\xi)$ as in the case of the continuous KN equation (i.e. 
the correspondence between the forms (\ref{eq:KN}) and 
(\ref{eq:KNrat}). We, thus, obtain for $\xi$ the equation: 
\begin{equation}\label{eq:ellBT}
-\xi_x\wt{\xi}_x=\frac{\sg(\xi+\wt{\xi}+\ar)\sg(\xi+\wt{\xi}-\ar)
\sg(\xi-\wt{\xi}+\ar)\sg(\xi-\wt{\xi}-\ar)}{\sg(2\ar)\sg(2\xi)
\sg(2\wt{\xi})}\  . 
\end{equation}
This elliptic form can be found from the original equation for the BT 
by employing a  number of relations, notably the identity: 
\begin{equation}\label{eq:Hrel}
H(u,v,a)=(u-v)^2 \left[\frac{1}{4}\left(\frac{U-V}{u-v}\right)^2-(u+v+a)\right] 
\left[\frac{1}{4}\left(\frac{U+V}{u-v}\right)^2-(u+v+a)\right] 
\end{equation}
in which $U^2\equiv r(u)$, $V^2\equiv r(v)$, as well as the addition formula for 
the Weierstrass $\wp$-function, i.e. 
\[ \wp(\xi)+\wp(\eta)+\wp(\xi+\eta)=\frac{1}{4}\left( \frac{\wp'(\xi)-\wp'(\eta)}
{\wp(\xi)-\wp(\eta)}\right)^2\    ,  \] 
and the relations 
\[\wp(\xi)-\wp(\eta)=\frac{\sg(\eta+\xi)\sg(\eta-\xi)}{\sg^2(\eta)
\sg^2(\xi)}~~~~, 
~~~~\wp'(\xi)=-\frac{\sg(2\xi)}{\sg^4(\xi)}\   , \] 
in which $\sg$ is the Weierstrass $\sg$-function.  
 
Eliminating the derivatives from the BT (\ref{eq:ellBT}) together with 
a second BT with $\wt{\phantom{a}}$  replaced with $\wh{\phantom{a}}$ 
and $\ar$ by $\bb$ we obtain
\begin{eqnarray} 
&&\frac{\sg(\xi+\wt{\xi}+\ar)\sg(\xi+\wt{\xi}-\ar)
\sg(\xi-\wt{\xi}+\ar)\sg(\xi-\wt{\xi}-\ar)}
{\sg(\xi+\wh{\xi}+\bb)\sg(\xi+\wh{\xi}-\bb)
\sg(\xi-\wh{\xi}+\bb)\sg(\xi-\wh{\xi}-\bb)}\times \nn\\ 
&&~~~ \times\frac{\sg(\wh{\xi}+\wh{\wt{\xi}}+\ar)\sg(\wh{\xi}+\wh{\wt{\xi}}-\ar)
\sg(\wh{\xi}-\wh{\wt{\xi}}+\ar)\sg(\wh{\xi}-\wh{\wt{\xi}}-\ar)}
{\sg(\wt{\xi}+\wh{\wt{\xi}}+\bb)\sg(\wt{\xi}+\wh{\wt{\xi}}-\bb)
\sg(\wt{\xi}-\wh{\wt{\xi}}+\bb)\sg(\wt{\xi}-\wh{\wt{\xi}}-\bb)}=
\frac{\sg^2(2\ar)}{\sg^2(2\bb)} \  . \label{eq:elldKN} 
\end{eqnarray} 
We remark that eq. (\ref{eq:elldKN}) cannot yet be regarded as 
the elliptic form of the Adler system (\ref{eq:dKN}), as the 
latter is obtained from a factorised form of the permutability 
condition of BT's. In fact, an elliptic version can be obtained 
in a straightforward way from the formula (\ref{eq:ellform}), but 
we haven't managed yet to cast it into an elegant form, so we omit 
it here. Alternatively, a so-called ``three-leg'' form of the Adler system was 
announced in the recent submission \cite{BS}, but it is yet unclear 
how this is related to the formula (\ref{eq:elldKN}).  
The purpose of these ways of rewriting the Adler system is to 
obtain a better insight into the underlying structure of 
the equation, including the origin of the Lax pair. 
A more systematic understanding of the relationship between 
the various members in the KN family of equations seems to be 
necessary and we intend to address these problems more at length in 
a future publication, \cite{HN}.

\subsection*{Acknowledgement}

The author acknowledges the hospitality of the Isaac Newton 
Institute of the University of Cambridge, where the present work was 
finalised. He is grateful for stimulating discussions with 
V. Papageorgiou, who brought the paper \cite{Adler} to his attention, 
and with J.Hietarinta, who performed the computations to check 
the consistency of the system (\ref{eq:dKN}) on the multidimensional 
lattice.

\end{document}